\documentclass[12pt]{article}  
\usepackage{cite}
\usepackage{epsfig}
\usepackage{graphicx}
\usepackage{amsmath}
\usepackage{amssymb}
\usepackage{ulem}
\usepackage{mdwlist}          
\usepackage{color}              

\usepackage{a41}
\usepackage{color}
\usepackage[rflt]{floatflt}
\usepackage{float}
\usepackage{slashed}

\newcommand{\lnMa}[1]{\ln^{#1}\biggl(\frac{m_1^2}{\mu^2}\biggr)}
\newcommand{\lnMb}[1]{\ln^{#1}\biggl(\frac{m_2^2}{\mu^2}\biggr)}

\usepackage{array}

\usepackage[english]{babel}
\usepackage[latin1]{inputenc}
\usepackage[T1]{fontenc}
\usepackage{ae}

\usepackage{url}

\usepackage{amsmath, amsthm, amssymb}
\newtheorem{thm}{Theorem}[section]

\newtheorem{definition}[thm]{Definition}

\usepackage{rotating}

\usepackage{graphicx}

\newcounter{mmacnt}
\def\restartmma{\setcounter{mmacnt}{0}}
\restartmma \catcode`|=\active
\def|#1|{\mathrm{#1}}
\catcode`|=12
\newenvironment{mma}{
 \par\smallskip
 \catcode`|=\active
 \parskip=0pt\parindent=0pt 
 \small
 \def\In##1\\{%
   \def\linebreak{\hfill\break\null\qquad}%
   \refstepcounter{mmacnt}
   \hangindent=2.5em\hangafter=0
   \leavevmode
   \llap{\tiny\sffamily In[\arabic{mmacnt}]:=\kern.5em}%
   \mathversion{bold}\footnotesize$\displaystyle##1$\normalsize
   \mathversion{normal}\par
 }%
 \def\Print##1\\{%
   \def\linebreak{\hfill\break}%
   \hangindent=2.5em\hangafter=0
   \leavevmode ##1\par}%
 \def\Out##1\\{%
   \def\linebreak{$\hfill\break\null\hfill$}%
   \kern\abovedisplayskip\par
   \hangindent=2.5em\hangafter=0
   \leavevmode
   \llap{\tiny\sffamily Out[\arabic{mmacnt}]=\kern.5em}
   \footnotesize$\displaystyle##1$\normalsize\hfill\null\par
   \kern\belowdisplayskip
 }%
 \def\Warning##1##2\\{%
   \def\linebreak{\hfill\break}%
   \hangindent=2.5em\hangafter=0
   \leavevmode
   {\scriptsize##1 : ##2}\par}%
}{%
 \par\smallskip
}

\usepackage{color}

\newenvironment{fshaded}{%
\MakeFramed {\FrameRestore}
}%
{\endMakeFramed}

\allowdisplaybreaks[4]

\begin{document}
\setlength{\baselineskip}{0.515cm}
\sloppy
\thispagestyle{empty}
\begin{flushleft}
DESY 17--157
\\
DO--TH 17/26\\
October  2017\\
\end{flushleft}

\mbox{}
\vspace*{\fill}
\begin{center}

{\LARGE\bf The Three Loop Two-Mass Contribution to the} 

\vspace*{3mm} 
{\LARGE\bf Gluon Vacuum Polarization}

\vspace{3cm}
\large
J.~Bl\"umlein$^a$, 
A.~De Freitas$^a$, 
C.~Schneider$^b$,
K.~Sch\"onwald$^a$

\vspace{1.cm}
\normalsize
{\it  $^a$ Deutsches Elektronen--Synchrotron, DESY,}\\
{\it  Platanenallee 6, D-15738 Zeuthen, Germany}
\\

\vspace*{3mm}
{\it $^b$~Research Institute for Symbolic Computation (RISC),\\
                          Johannes Kepler University, Altenbergerstra\ss{}e 69,
                          A--4040, Linz, Austria}\\


\end{center}
\normalsize
\vspace{\fill}
\begin{abstract}
\noindent
We calculate the two-mass contribution to the 3-loop vacuum polarization of the gluon in 
Quantum Chromodynamics at virtuality $p^2 = 0$ for general masses and also present the 
analogous result for the photon in Quantum Electrodynamics. 
\end{abstract}

\vspace*{\fill}
\noindent

\newpage
The 3-loop heavy flavor corrections to deep-inelastic scattering at larger virtualities \cite{
Bierenbaum:2009mv,Behring:2014eya,Ablinger:2010ty,Ablinger:2014vwa,Ablinger:2014nga,
Ablinger:2014lka,Ablinger:2016swq} form an important ingredient for the determination of the strong 
coupling constant $\alpha_s(M_Z^2)$, the parton distribution functions and the measurement of the mass 
of the charm quark $m_c$ at high precision \cite{Alekhin:2012vu,Alekhin:2017kpj}. Starting at 3-loop 
order the QCD corrections contain also 2-mass contributions in single Feynman diagrams 
\cite{Ablinger:2017err,Ablinger:2011pb,Ablinger:2012qj}. The heavy flavor contributions 
to deep-inelastic structure functions in the region of larger virtualities $Q^2 \gg m_Q^2$, with $m_Q$ the 
heavy quark mass, can be obtained in terms of massive on-shell operator matrix elements (OMEs) 
\cite{Buza:1995ie,Bierenbaum:2009mv}. These quantities also receive massive self-energy insertions, such as the
on-shell vacuum polarization function
$\hat{\tilde{\Pi}}^{(3)}(0,m_1^2,m_2^2,\mu^2)$ and fermion self energy $\hat{\tilde{\Sigma}}(0,m_1^2,m_2^2,
\mu^2)$. Here $m_{1,2}$ denote the corresponding heavy quark masses and $\mu$ is the renormalization scale.
These quantities are of more general interest, as they appear in various massive higher loop calculations.
The expression for  $\hat{\tilde{\Sigma}}(0,m_1^2,m_2^2,\mu^2)$ for general ratios $\eta = m_1^2/m_2^2$
has been given in Ref.~\cite{CS1,Ablinger:2017err}. In the present note, we calculate the polarization function
$\hat{\tilde{\Pi}}^{(3)}(0,m_1^2,m_2^2,\mu^2)$, which is obtained as the 3rd term in the expansion
\begin{eqnarray}
\hat{\Pi}^{\mu\nu}_{ab,H}(p^2,m_1^2,m_2^2,\mu^2,\varepsilon,\hat{a}_s) 
= i (-p^2 g^{\mu\nu} + p^\mu p^\nu) \delta_{ab} \sum_{k=1}^\infty \hat{a}_s^k 
\hat{\Pi}_{H}(p^2,m_1^2,m_2^2,\mu^2,\varepsilon)   
\end{eqnarray}
in the limit $p^2 \rightarrow 0$. Here the index $H$ labels the heavy quark part of the polarization function,
$\hat{a}_s = g_s^2/(4\pi)^2$ denotes the unrenormalized strong coupling constant, 
$m_{1,2}$ are the bare heavy quark masses, and $\varepsilon = D - 4$ is the dimensional parameter. In the calculation
we refer to the Feynman rules given in \cite{YND}. 
The corresponding (single mass) expressions 
$\hat{{\Pi}}^{(1,2)}_H(0,m_1^2,m_2^2,\mu^2)$ were calculated in \cite{Kallen:1955fb,Broadhurst:1991fi,
Fleischer:1992re,Jegerlehner:1998zg,Chetyrkin:2008jk,Bierenbaum:2009mv} for QED and/or QCD.

There are six physical topologies contributing to  $\hat{\tilde{\Pi}}^{(3)}(0,m_1^2,m_2^2,\mu^2)$, Eq.~(\ref{eq:pi2m}).
The corresponding 3-loop Feynman diagrams have been generated using {\tt QGRAF} \cite{Nogueira:1991ex}
and the color-traces were calculated using {\tt Color} \cite{vanRitbergen:1998pn}. Standard Feynman 
parameter integration has been applied, cf.~\cite{Ablinger:2015tua}, representing one of the 
integrals using the Mellin-Barnes contour integral \cite{MB1a,MB1b,MB2,MB3,MB4}, also using the package
\cite{Czakon:2005rk}. The sums over the residues have been performed analytically 
using the packages {\tt Sigma} \cite{SIG1,SIG2}, {\tt EvaluateMultiSums}
and {\tt SumProduction} \cite{EMSSP}, applying procedures of {\tt HarmonicSums} 
\cite{HARMONICSUMS,Ablinger:PhDThesis,Ablinger:2011te,Ablinger:2013cf,Ablinger:2014bra}  also for the limiting 
processes in case of infinite sums.

The equal mass contributions have been calculated in 
\cite{Bierenbaum:2009mv}, Eq.~(4.7), before in case of a general 
$R_\xi$ gauge using {\tt MATAD} \cite{Steinhauser:2000ry}. The 2-mass term is given by 
\begin{eqnarray}
\hat{\tilde{\Pi}}^{(3)}(0,m_1^2,m_2^2,\mu^2) &=& \lim_{p^2\rightarrow 0}
\hat{\Pi}_{H}(p^2,m_1^2,m_2^2,\mu^2,\varepsilon)
\nonumber\\ &=& 
-C_F T_F^2 \Biggl\{
\frac{256}{9 \varepsilon^2}
+\frac{64}{3 \varepsilon} \left[\lnMa{}+\lnMb{}+\frac{5}{9}\right]
-5 \eta -\frac{5}{\eta}
\nonumber \\ &&
+\left(-\frac{5 \eta }{8}-\frac{5}{8 \eta }+\frac{51}{4}\right) \ln ^2(\eta )+\left(\frac{5}{2 \eta }-\frac{5 \eta }{2}\right) \ln (\eta)
+\frac{32 \zeta_2}{3}
\nonumber \\ &&
+32 \lnMa{} \lnMb{}
+\frac{80}{9} \lnMa{}+\frac{80}{9} \lnMb{}
+\frac{1246}{81}
\nonumber \\ &&
+\left(\frac{5 \eta^{3/2}}{2}
+\frac{5}{2 \eta^{3/2}}
+\frac{3 \sqrt{\eta}}{2}
+\frac{3}{2 \sqrt{\eta}}\right) \Biggl[\frac{1}{8} \ln \left(\frac{1+\sqrt{\eta}}{1-\sqrt{\eta}}\right) \ln^2(\eta)
\nonumber \\ &&
-{\rm Li}_3\left(-\sqrt{\eta}\right)
+{\rm Li}_3\left(\sqrt{\eta}\right)
-\frac{1}{2} \ln (\eta) \left({\rm Li}_2\left(\sqrt{\eta}\right)
-{\rm Li}_2\left(-\sqrt{\eta}\right)\right)\Biggr]
\Biggr\}
\nonumber \\ &&
-C_A T_F^2 \Biggl\{
\frac{64}{9 \varepsilon^3} \left(1+2 \xi\right)
+\frac{16}{3 \varepsilon^2} \Biggl[(1+2 \xi) \left(\lnMa{}+\lnMb{}\right)
\nonumber \\ &&
-\frac{35}{9}\Biggr]
+\frac{4}{\varepsilon} \Biggl[
\lnMa{2}+\lnMb{2}
-\frac{35}{9} \lnMa{}
-\frac{35}{9} \lnMb{}
\nonumber \\ &&
+\frac{2}{3} \zeta_2
+\frac{37}{27}
+\xi \left(\frac{4}{3} \ln^2(\eta)+4 \lnMa{} \lnMb{}+\frac{4}{3} \zeta_2+\frac{292}{81}\right)
\Biggr]
\nonumber \\ &&
+2 \left(1+\xi\right) \left(\lnMa{3}+\lnMb{3}\right)
-\frac{70}{3} \lnMa{} \lnMb{}
\nonumber \\ &&
+2 \xi \lnMa{2} \lnMb{}
+2 \xi \lnMa{} \lnMb{2}
-\frac{2}{9} \left(2+\xi\right) \ln^3(\eta)
\nonumber \\ &&
+\left(2 \left(1+2 \xi\right) \zeta_2+\frac{37}{9}+\frac{292}{27} \xi\right) \left(\lnMa{}+\lnMb{}\right)
\nonumber \\ &&
+\left[\frac{4}{3} \left(2+\xi\right) \ln(1-\eta)-\left(\eta+\frac{1}{\eta}\right) \left(\frac{2}{3}+\frac{5 \xi}{24}\right)-\frac{179}{18}-\frac{43}{36} \xi\right] \ln^2(\eta)
\nonumber \\ &&
-\frac{1}{3} \left(16+5 \xi\right) \left(\eta+\frac{1}{\eta}\right)
-\frac{70}{9} \zeta_2
-\frac{8}{9} \zeta_3 \left(7+2 \xi\right)
-\frac{3769}{243}
+\frac{262}{243} \xi
\nonumber \\ &&
+\left(\frac{1}{\eta}-\eta\right) \left(\frac{8}{3}+\frac{5}{6} \xi\right) \ln(\eta)
+\frac{8}{3} \left(2+\xi\right) \big({\rm Li}_2(\eta) \ln(\eta)-{\rm Li}_3(\eta)\big)
\nonumber \\ &&
+\left[\left(8+\frac{5}{2} \xi\right) \frac{1+\eta^3}{3 \eta^{3/2}}+\left(10+\frac{9}{2} \xi\right) \frac{1+\eta}{\sqrt{\eta}}\right]
\Biggl[\frac{1}{8} \ln \left(\frac{1+\sqrt{\eta}}{1-\sqrt{\eta }}\right) \ln^2(\eta)
\nonumber \\ &&
-{\rm Li}_3\left(-\sqrt{\eta}\right)
+{\rm Li}_3\left(\sqrt{\eta}\right)
-\frac{1}{2} \ln(\eta) \left({\rm Li}_2\left(\sqrt{\eta}\right)-{\rm Li}_2\left(-\sqrt{\eta}\right)\right)\Biggr]
\Biggr\}.
\label{eq:pi2m}
\end{eqnarray}
Here $C_A = N_c, C_F = (N_c^2-1)/(2 N_c), T_F = 1/2$
are the color factors and $N_c = 3$ in case of QCD.

Using {\tt Q2E}/{\tt Exp} \cite{Harlander:1997zb,Seidensticker:1999bb} the first terms of the 
$\eta$-expansion of (\ref{eq:pi2m}) have been obtained in Ref.~\cite{Ablinger:2017err} before. The corresponding 
expansion of (\ref{eq:pi2m}) agrees with this expression. Eq.~(\ref{eq:pi2m}) is symmetric under the
interchange $m_1 \Leftrightarrow m_2$ and depends on the gauge parameter $\xi$ at most linearly. Although
not explicitly visible in the representation given above, one can show that 
$\hat{\tilde{\Pi}}^{(3)}(0,m_1^2,m_2^2,\mu^2)$ depends only on $\eta$ and not on $\sqrt{\eta}$.

The corresponding expression in case of Quantum Electrodynamics is found by setting $C_F = 1, C_A = 0$ and
$T_F = 1$. From \cite{Lee:2013sx} the 3-loop QED result can be inferred in principle.

\noindent
{\bf Acknowledgment.}~
We would like to thank P.~Marquard for discussions and M.~Steinhauser for providing the codes {\tt MATAD 3.0} and 
{\tt Q2E}/{\tt Exp}. Discussions with A.~Behring are gratefully acknowledged. This work was supported in part by the European 
Commission through PITN-GA-2012-316704 ({HIGGSTOOLS}).

{\small
}
\end{document}